\documentclass[12pt]{article}
\usepackage{amsmath}
\usepackage{amssymb}
\usepackage[english]{babel}
\usepackage{latexsym}
\usepackage{times}
\usepackage{mathptm}
\begin{document}
\small
\normalsize
\newcounter{saveeqn}
\newcommand{\alpheqn}{\setcounter{saveeqn}{\value{equation}}%
\setcounter{equation}{0}%
\renewcommand{\theequation}{\mbox{\arabic{saveeqn}-\alph{equation}}}}
\newcommand{\reseteqn}{\setcounter{equation}{\value{saveeqn}}%
\renewcommand{\theequation}{\arabic{equation}}}
\renewcommand{\thesection}{\Roman{section}}

\protect\newtheorem{principle}{Principle} 
\protect\newtheorem{theo}[principle]{Theorem}
\protect\newtheorem{prop}[principle]{Proposition}
\protect\newtheorem{lem}[principle]{Lemma}
\protect\newtheorem{co}[principle]{Corollary}
\protect\newtheorem{de}[principle]{Definition}
\newtheorem{ex}[principle]{Example}
\newtheorem{rema}[principle]{Remark}
\newtheorem{state}[principle]{Statement}{\bf}{\rm}
\newtheorem{proof}{Proof}{\it}{\rm}
\newtheorem{acknowledgements}[principle]{Acknowledgements}{\bf}{\rm}
\small
\normalsize
\noindent {\Large \textsf{A new class of entanglement measures} } \\ \\
{\large {Oliver Rudolph}}  \\ \vskip.0001in
\noindent {\normalsize
Physics Division, Starlab nv/sa, Engelandstraat 555, B-1180 Brussels,
Belgium.} \\
{\normalsize Email: o.rudolph@ic.ac.uk}
\\ \\
\normalsize
\noindent \emph{Abstract} We introduce new entanglement measures
on the set of
density operators on tensor product Hilbert spaces. These
measures are based on the greatest cross norm on the
tensor product of the sets of trace class operators on Hilbert
space.
We show that they satisfy the
basic requirements on entanglement measures
discussed in the literature, including convexity, invariance under
local unitary operations and non-increase under local quantum
operations and classical communication.
\section{Introduction}
This paper is devoted to the study of
entanglement of quantum states, which is one of the most decisively
non-classical features in quantum theory.
The question of quantifying entanglement in the case of mixed
quantum states represented by
density operators on finite dimensional
Hilbert spaces has recently been studied extensively
in the context of quantum information
theory, see, e.g.,
\cite{Rudolph00,BennettVSW97,VedralPRK97,PlenioV98,VedralP98,Horodecki99,Vidal00,Horodecki98,BennettVMSST98,Rudolph01}
and references therein.

An \emph{entanglement measure} is a real-valued function defined
on the set of density operators on some tensor product Hilbert
space subject to further physically motivated conditions, see,
e.g., \cite{VedralPRK97,PlenioV98,VedralP98,Horodecki99,Vidal00}
and below. A number of entanglement measures have been discussed in the
literature, such as the von Neumann reduced entropy, the relative
entropy of entanglement \cite{PlenioV98},
the entanglement of distillation
and the entanglement of formation \cite{BennettVSW97}.
Several authors proposed physically motivated postulates
to characterize entanglement
measures, see below. These postulates (although they vary from author to
author in the details) have in common that they are
based on the concepts of the operational formulation of quantum mechanics
\cite{Kraus83}. We shall discuss one version of these
\emph{operational characterizations} of entanglement measures in
Section IV.

In this paper we introduce new entanglement
measures based on the greatest cross norm on the tensor product of
the sets of trace class operators on Hilbert space (see Sections V and VI).
We shall show
that the measures introduced in this work satisfy
all the basic requirements
for entanglement measures. These
include convexity, invariance under local unitary
transformations, and non-increase under procedures composed
of local quantum operations and classical communication.

Throughout this paper the set of trace class operators on some Hilbert
space ${\cal H}$ is denoted by ${\cal T}({\cal H})$ and
the set of bounded operators on ${\cal H}$ by
${\cal B}({\cal H})$. A density operator is a positive trace
class operator with trace one.

\section{Preliminaries}
In this section we collect some basic definitions and results
which are used in the course of this paper.

In the present paper we restrict ourselves mainly to the situation of
a composite quantum system consisting of two subsystems with Hilbert space
${\mathcal{H}}_1 \otimes {\mathcal{H}}_2$ where ${\mathcal{H}}_1$ and
${\mathcal{H}}_2$ denote the Hilbert spaces of the subsystems
(except in Section VI).
The states of the system are identified with the density operators
on ${\mathcal{H}}_1 \otimes {\mathcal{H}}_2$.
\begin{de}
Let ${\cal H}_1$ and ${\cal H}_2$ be two Hilbert spaces of
arbitrary dimension. A density operator $\varrho$ on the tensor product
${\cal H}_1
\otimes {\cal H}_2$ is called \emph{separable} or
\emph{disentangled} if there exist a
family $\left\{ \omega_{i} \right\}$ of positive real numbers, a family
$\left\{ \rho^{(1)}_i \right\}$ of density operators on
${\cal H}_1$ and a family $\left\{ \rho^{(2)}_i \right\}$ of
density operators
on ${\cal H}_2$ such that
\begin{equation} \label{e1}
\varrho = \sum_{i} \omega_{i} \rho^{(1)}_i \otimes \rho^{(2)}_i,
\end{equation}
where the sum converges in trace class norm.
\end{de}
The set of states is a convex set and its extreme points,
which are also called \emph{pure states}, are the
projection operators. Every pure state obviously correspond to a
unit vector $\psi$ in ${\mathcal{H}}_1 \otimes {\mathcal{H}}_2$.
We denote the projection operator onto the subspace spanned by the
unit vector $\psi$ by $P_\psi$.

The Schmidt decomposition is of central importance in the
characterization and quantification of entanglement associated
with pure states.
\begin{lem}
Let ${\mathcal{H}}_1$ and ${\mathcal{H}}_2$ be Hilbert spaces of
arbitrary dimension and let $\psi \in {\mathcal{H}}_1 \otimes
{\mathcal{H}}_2$.
Then there exist a family of non-negative real numbers $\{
p_i \}_i$ and orthonormal bases $\{ a_i \}_i$ and $\{ b_i \}_i$ of
${\mathcal{H}}_1$ and ${\mathcal{H}}_2$ respectively such that
\[ \psi = \sum_i \sqrt{p_i} a_i \otimes b_i. \] \label{Sch}
\end{lem}
The family of positive numbers $\{ p_i \}_i$ is called the family
of \emph{Schmidt coefficients} of $\psi$.
For pure states the family of Schmidt coefficients of a state
completely characterizes the amount of entanglement of that state.
A pure state $\psi$ is separable if and only if $\psi = a \otimes
b$ for some $a \in {\mathcal{H}}_1$ and $b \in {\mathcal{H}}_2$.

The \emph{von Neumann reduced entropy}
for density operators
$\sigma$ on a tensor product Hilbert space ${\mathcal{H}}_1 \otimes
{\mathcal{H}}_2$ is defined as
\begin{equation}
S_{\mathrm{vN}}(\sigma) := -
{\mathrm{Tr}}_{{\mathcal{H}}_1}({\mathrm{Tr}}_{{\mathcal{H}}_2}
\sigma \ln ({\mathrm{Tr}}_{{\mathcal{H}}_2}
\sigma)),
\end{equation} where
${\mathrm{Tr}}_{{\mathcal{H}}_1}$ and ${\mathrm{Tr}}_{{\mathcal{H}}_2}$
denote the partial traces over
${\mathcal{H}}_1$ and ${\mathcal{H}}_2$ respectively. In the case of
pure states $\sigma = P_\psi$, it can be shown that
$- {\mathrm{Tr}}_{{\mathcal{H}}_1}({\mathrm{Tr}}_{{\mathcal{H}}_2}
P_\psi
\ln ({\mathrm{Tr}}_{{\mathcal{H}}_2} P_\psi)) = $ \\ $-
{\mathrm{Tr}}_{{\mathcal{H}}_2}({\mathrm{Tr}}_{{\mathcal{H}}_1} P_\psi
\ln ({\mathrm{Tr}}_{{\mathcal{H}}_1} P_\psi)) = - \sum_i p_i \ln p_i$
where $\{ p_i \}_i$ denotes the family of Schmidt coefficients of $\psi$.
However, for a general mixed state $\sigma$ we have
${\mathrm{Tr}}_{{\mathcal{H}}_1}({\mathrm{Tr}}_{{\mathcal{H}}_2} \sigma
\ln ({\mathrm{Tr}}_{{\mathcal{H}}_2} \sigma)) \linebreak[3] \neq
{\mathrm{Tr}}_{{\mathcal{H}}_2}({\mathrm{Tr}}_{{\mathcal{H}}_1} \sigma
\ln ({\mathrm{Tr}}_{{\mathcal{H}}_1} \sigma))$.

\section{Effects and Operations}
In this section we recall some of the fundamental concepts and definitions
in the operational approach to quantum theory and in
quantum measurement theory
\cite{Kraus83,Davies76,Kraus71,BuschGL95,BuschLM96}.
The quantum mechanical \emph{state} of a quantum system is described by a
density operator $\varrho$ on the system's Hilbert space $\mathcal{H}$,
i.e., by a positive trace class operator with trace one.
Let $\mathcal{K}$ be another Hilbert space.
An \emph{operation} is a positive linear map $T : \mathcal{T}(\mathcal{H})
\to \mathcal{T}(\mathcal{K})$ such that $T$ is trace non-increasing
for positive trace class operators,  i.e., $0 \leq {\mathrm{Tr}}(T (\sigma))
\leq {\mathrm{Tr}}(\sigma)$ for all positive $\sigma \in
\mathcal{T}(\mathcal{H}).$ Following \cite{BennettVSW97,Vidal00} we
adopt the point of view that allowed operations in a laboratory are
(O1) adding an ancilla, (O2) tracing out
part of the system, (O3) performing unitary operations, and (O4)
performing possibly selective yes-no experiments.
It can be shown (for a detailed proof see, e.g., \cite{DonaldHR01})
that the class of operations $T : \mathcal{T}(\mathcal{H})
\to \mathcal{T}(\mathcal{K})$ composed out of operations of the
form (O1)-(O4) coincides with the class of trace non-increasing
\emph{completely positive} operations, i.e.,
has the property that for all $n \geq 0$ the map $T_n$ on
${\mathcal{T}}({\mathcal{H}} \otimes {\mathbb{C}}^n)$ defined by
$T_n := T \otimes 1_n$, where
$1_n \in {\mathcal{B}}({\mathbb{C}}^n)$ denotes the unit matrix,
is positive. For a further physical motivation of the requirement of
complete positivity see, e.g.,
\cite{Kraus83}.
In the sequel it is always understood that all operations are
completely positive. If ${\mathcal{H}}$ and ${\mathcal{K}}$ are
both finite dimensional Hilbert spaces, then it
follows from the Choi-Kraus
representation theorem for operations
\cite{Kraus83,Kraus71,Choi75} that for
every operation $T : {\mathcal{T}}({\mathcal{H}}) \to
{\mathcal{T}}({\mathcal{K}})$ there exists a
family of bounded operators $\{ A_k : {\mathcal{H}} \to {\mathcal{K}}
\}_k$ with $\sum_k A_k A_k^\dagger \leq 1_{{\mathcal{K}}}$
such that $T$ can be expressed as
\begin{equation}
T(\sigma) = \sum_k A_k^\dagger \sigma A_k \label{eq1} \end{equation}
for all $\sigma \in
{\mathcal{T}}({\mathcal{H}})$.
If ${\mathcal{H}} = {\mathcal{K}}$, the Choi-Kraus representation
is also valid for infinite dimensional Hilbert spaces (all
sums converge in trace class norm).
The family $\{ A_k \}_k$ is not unique.
However, the operator $E := \sum_k A_k A_k^\dagger = T^*(1)$
is independent of the
family $\{ A_k \}_k$ chosen and is called the \emph{effect}
corresponding to the operation $T$ and its associated yes-no
measurement ($T^*$ denotes the adjoint of $T$, \cite{Kraus83}).
Generally, an operator $E$ is called an \emph{effect operator}
if $E$ is bounded and Hermitean and if $0 \leq E \leq 1$.
\emph{Effect operator valued measures}
are then the most general \emph{observables} in the theory \cite{BuschGL95}.
They are also called \emph{positive operator valued (POV)
measures}. A \emph{L\"uders-von Neumann
operation} is  an operation of the form
$T_L(\sigma) = \sum_k P_k \sigma P_k$ where $\{ P_k \}_k$
is a set of mutually orthogonal projection operators
on ${\cal H}$.
L\"uders-von Neumann operations are repeatable.
In the case of a general operation, it is possible to view the
terms in its Choi-Kraus representation as representing different
possible measurement outcomes. In the terminology of operational quantum
theory the individual terms in the Choi-Kraus representation (\ref{eq1}) form a
set of operations corresponding to coexistent effects, see
\cite{Kraus83,BuschGL95}:
two effect operators $E_1$ and $E_2$ are
called \emph{coexistent} if
there exist effect operators $F,G,H$ with $F+G+H \leq 1$
such that $E_1 = F + G$ and $E_2 = F+H$ (in general $F,G$ and $H$
will not be unique however).
Therefore in general two coexistent effect operators $E_1$ and $E_2$
do not correspond to mutually complementary measurement outcomes
but instead may have some `overlap' represented by the operator $F$
even if $E_1 + E_2 \leq 1$. Coexistent effect operators need not
commute.
\section{Entanglement measures}
An entanglement measure is a functional $E$ defined
on the set of density operators on the Hilbert space of a
composite quantum system measuring the degree
of entanglement of every given density operator.
Every measure
of entanglement $E$ should satisfy the following requirements
\cite{BennettVSW97,VedralPRK97,PlenioV98,Horodecki99,Vidal00}
\begin{itemize}
\item[(E0)] An entanglement measure is a positive real-valued
functional
$E$ which for any given two systems is well-defined
on the set ${\cal D}({\cal H}_1 \otimes {\cal H}_2)$
of density operators on the
tensor product ${\cal H}_1 \otimes {\cal H}_2$
of the Hilbert spaces of the
two systems. Moreover, $E$ is \emph{expansible}, i.e.,
whenever $\rho \in {\cal D}({{\mathcal{H}}_1} \otimes
{{\mathcal{H}}_2}) \subset {\cal D}({\mathsf{H}}_1 \otimes
{\mathsf{H}}_2)$ with embeddings
${{\mathcal{H}}_1} \hookrightarrow {\mathsf{H}}_1$ and
${{\mathcal{H}}_2} \hookrightarrow {\mathsf{H}}_2$
of Hilbert spaces
${{\mathcal{H}}_1}$ and ${{\mathcal{H}}_2}$ into larger Hilbert spaces
${\mathsf{H}}_1$ and ${\mathsf{H}}_2$ respectively, then
$E \vert_{{{\mathcal{H}}_1} \otimes {{\mathcal{H}}_2}}(\rho) =
E \vert_{{\mathsf{H}}_1 \otimes {\mathsf{H}}_2}(\rho)$.
\item[(E1)] If $\sigma$ is separable, then $E(\sigma) = 0$.
\item[(E2)] Local unitary transformations
leave $E$ invariant, i.e., \[ E(\sigma) =
E \left((U_1 \otimes U_2) \sigma (U_1^\dagger \otimes U_2^\dagger) \right)
\]
for all unitary operators $U_1$ and $U_2$ acting on ${\cal H}_1$ or
${\cal H}_2$ respectively.
\item[(E3)] Entanglement cannot increase under procedures
consisting of
local operations on the two quantum systems and classical
communication. If $T$ is an operation which is trace-preserving
on positive operators and can be realized by
means of local operations and classical communication,
i.e., is composed out of local operations of the form (O1) - (O4)
and classical communication, then
\addtocounter{equation}{1}
\alpheqn
\begin{equation}
E(T(\sigma)) \leq E(\sigma) \label{locc}
\end{equation} for all $\sigma \in \mathcal{D}({\mathcal{H}}_1
\otimes {\mathcal{H}}_2)$.
It is clear that every procedure acting on an individual single
quantum system
${\mathcal{H}}_1 \otimes {\mathcal{H}}_2$ composed only of local
operations and classical communication can
formally be represented as a finite sequence of operations
of the form $T_1 \otimes T_2$, where $T_1$ and $T_2$ are
local operations on ${\mathcal{H}}_1$ and ${\mathcal{H}}_2$
respectively. The requirement that entanglement cannot increase
under local operations and classical communication
is thus equivalent to
\begin{equation} \label{4b}
E((T_1 \otimes T_2) (\sigma)) \leq E(\sigma),
\end{equation} \reseteqn
for all $\sigma \in \mathcal{D}({\mathcal{H}}_1
\otimes {\mathcal{H}}_2)$.
\end{itemize}
\begin{rema} Equation (\ref{4b}) stipulates that local
operations cannot increase entanglement.
In the quantum information literature most authors
replace Equations (\ref{locc}) and (\ref{4b}) by
the stronger requirement
\begin{equation} \label{blabla}
\sum_i p_i E(\sigma_i) \leq E(\sigma). \end{equation}
Equation \ref{blabla} stipulates that after the measurement
the entanglement (as measured by
$E$) averaged over the possible output states $\sigma_i$ is less
than or equal to the original entanglement. Here $p_i$ denotes the probability
that the final state $\sigma_i$ occurs. In the literature
Equation \ref{blabla}
is normally taken as the formal expression for the paradigm that
it is impossible to create or increase entanglement by performing
procedures composed of local quantum operations and
classical communication alone. A disadvantage of Equation (\ref{blabla})
is that it makes sense only in measurement situations and that
the `possible output states' corresponding to a given operation
$T$ are not uniquely defined. Mathematically this corresponds to the
fact that the Choi-Kraus representation of an operation $T$ is in general
not unique. The difference between Equations
(\ref{locc}) and (\ref{blabla}) is that Equation (\ref{blabla})
stipulates that entanglement cannot increase \emph{on average}
under local operations and classical communications
(for a detailed discussion of this point see \cite{Vidal00}).
In contrast Equation (\ref{locc}) says that entanglement
cannot increase for any
operation which acts on individual systems
and is composed of local operations and classical
communication. If one takes up the former (ensemble) point of view of
Equation (\ref{blabla}), then Equation (\ref{4b}) does no longer
represent the most general condition because
from the ensemble point of view the most general operations composed out
of local operations and classical communications can contain
correlations between terms of the Choi-Kraus representations of
subsequent local operations. A precise definition can be found, e.g., in
\cite{DonaldHR01}. Some authors consider also other classes of local
operations, most prominently the class of \emph{separable} operations
considered in \cite{VedralPRK97}.
\end{rema}
\begin{itemize}
\item[(E4)] Mixing of states does not increase entanglement, i.e.,
$E$ is convex \[ E(\lambda \sigma + (1 - \lambda) \tau) \leq
\lambda E(\sigma) + (1 - \lambda) E(\tau) \] for all $0 \leq
\lambda \leq 1$ and all $\sigma, \tau \in \mathcal{D}({\mathcal{H}}_1
\otimes {\mathcal{H}}_2)$.
\end{itemize}
Apart from the requirements (E0) - (E4) on entanglement measures
many authors add further requirements to the definition of
entanglement measures but we are not going to discuss them in this
paper. For a details, see \cite{DonaldHR01}.
In the sequel we exclude the trivial
functional $E \equiv 0$ which also satisfies (E0) - (E4).
\begin{rema}
Postulate (E2) is an immediate consequence of (E3).
\end{rema}
\begin{rema} It has been argued in \cite{Horodecki98} that
the entanglement of distillation $E_D$ introduced in \cite{BennettVSW97}
does vanish for certain
non-separable states (so called bound entangled states). Therefore it
has been pointed out in \cite{Horodecki99} that replacing (E1)
by the stronger requirement that for every entanglement measure
$E(\sigma) =0$ if and only if $\sigma$ is separable might exclude
interesting entanglement measures. For more information the
reader is referred to
the references. \end{rema}
\begin{ex} \label{ex1}
Post selection of a subensemble means selecting a
(non-normalized) output state of a
quantum operation and normalizing its trace to 1.
This procedure can lead to an increase in entanglement. This can
be seen by considering a very simple example. Consider a composite
quantum system composed of two
3-level quantum systems and the state \[ \rho_\epsilon = (1 - \epsilon)
\vert 0 0 \rangle \langle 00 \vert + \frac{\epsilon}{2}  (\vert 1 2 \rangle -
\vert 21 \rangle ) ( \langle 12 \vert - \langle 21 \vert ). \]
For $\epsilon$ small it is intuitively obvious that
this state does not contain `much' entanglement and every
entanglement measure should reflect this. Indeed, consider for example the
relative entropy of entanglement introduced in \cite{VedralPRK97}
defined by \begin{equation} E_S(\sigma) :=
\inf_\rho \left( {\mathrm{Tr}} (
\sigma \ln \sigma - \sigma \ln \rho ) \right) \label{relent} \end{equation}
where the infimum runs over all separable states $\rho$ for which
${\mathrm{Tr}}(\sigma \ln \rho)$ is well-defined and finite.
Elementary estimates using the results of \cite{VedralPRK97}
show that \[ E_S(\rho_\epsilon) \leq \epsilon
\ln 2. \] If we subject the system to an operation testing whether
or not
the system is in the state $\vert 00 \rangle$ and select after the
measurement the subensemble corresponding to the negative outcome (system
is not in the state $\vert 00 \rangle$), then clearly the final
state after the operation and post selection is given by
$\frac{1}{2}  (\vert 1 2 \rangle -
\vert 21 \rangle ) ( \langle 12 \vert - \langle 21 \vert )$.
Notice that this operation can be achieved by local operations and
classical communication. We
find \[ E_S \left( \frac{1}{2}  (\vert 1 2 \rangle -
\vert 21 \rangle ) ( \langle 12 \vert - \langle 21 \vert ) \right)
= \ln 2 > E_S(\rho_\epsilon). \]
Similarly it can be shown that the entanglement measure
$\Vert \cdot \Vert_\gamma$ to be
introduced below may increase under post selection of
subensembles. Therefore we see that we must not replace the
operation in Equation (\ref{locc}) by some normalized
non-linear operation
$\rho \mapsto \frac{T(\rho)}{{\mathrm{Tr}}(T(\rho))}$
corresponding to post selection of a subensemble.
\end{ex}
\section{A new class of entanglement measures}
Consider the situation that the two Hilbert spaces ${\cal H}_1$ and ${\cal
H}_2$ are both finite dimensional and
consider the spaces ${\cal T}({\cal H}_1)$ and ${\cal T}({\cal H}_2)$
of trace class operators on ${\cal H}_1$ and ${\cal H}_2$
respectively. Both spaces are Banach spaces when equipped with the trace
class norm $\Vert \cdot \Vert_1^{(1)}$ or $\Vert \cdot
\Vert_1^{(2)}$ respectively, see, e.g., Schatten \cite{Schatten70}. In
the sequel we shall drop the superscript and write $\Vert \cdot
\Vert_1$ for both norms, slightly abusing the notation; it will
always be clear from the context which norm is meant.
The algebraic tensor product ${\cal T}({\cal H}_1)
\otimes_{\rm alg} {\cal T}({\cal H}_2)$
of ${\cal T}({\cal H}_1)$
and ${\cal T}({\cal H}_2)$ is defined as the set of all finite
linear combinations of elementary tensors $u \otimes
{v}$, i.e., the set of all finite sums $\sum_{i=1}^n u_i \otimes
{v}_i$ where $u_i \in {\cal T}({\cal H}_1)$ and
${v}_i \in {\cal T}({\cal H}_2)$ for all $i$.

It is well known that we can define a cross norm on ${\cal T}({\cal H}_1)
\otimes_{\rm alg} {\cal T}({\cal H}_2)$ by \cite{WeggeOlsen93}
\begin{equation} \label{gamma}
\Vert t \Vert_\gamma := \inf \left\{ \sum_{i=1}^n
\left\Vert u_i \right\Vert_1 \, \left\Vert
{v}_i \right\Vert_1 \, \left\vert \, t = \sum_{i=1}^n u_i
\otimes {v}_i \right. \right\}, \end{equation} where $t \in
{\cal T}({\cal H}_1)
\otimes_{\rm alg} {\cal T}({\cal H}_2)$ and where the infimum runs over
all finite decompositions of $t$ into elementary tensors.
It is also well known that $\Vert \cdot \Vert_\gamma$ majorizes any
subcross seminorm on ${\cal T}({\cal H}_1)
\otimes_{\rm alg} {\cal T}({\cal H}_2)$. We denote the completion
of ${\cal T}({\cal H}_1)
\otimes_{\rm alg} {\cal T}({\cal H}_2)$ with respect to
$\Vert \cdot \Vert_\gamma$ by ${\cal T}({\cal H}_1)
\otimes_{\gamma} {\cal T}({\cal H}_2)$. ${\cal T}({\cal H}_1)
\otimes_{\gamma} {\cal T}({\cal H}_2)$
is a Banach algebra \cite{WeggeOlsen93}.

As both ${\cal H}_1$ and ${\cal
H}_2$ are finite dimensional, ${\cal T}({\cal H}_1) =
{\cal B}({\cal H}_1)$ and ${\cal T}({\cal H}_2) =
{\cal B}({\cal H}_2)$ and ${\cal B}({\cal H}_1)
\otimes_{\rm alg} {\cal B}({\cal H}_2) =
{\cal B}({\cal H}_1 \otimes {\cal H}_2)$, see,
e.g., \cite{KadisonR86}, Example 11.1.6. In finite
dimensions all Banach space norms on ${\cal B}({\cal H}_1
\otimes {\cal H}_2)$, in particular
the operator norm $\Vert \cdot \Vert$, the trace class norm
$\Vert \cdot \Vert_1$, and the norm $\Vert \cdot
\Vert_\gamma$, are
equivalent, i.e., generate the same topology on ${\cal
B}({\cal H}_1 \otimes {\cal H}_2)$.

For later reference we compute the value of $\Vert P_\psi \Vert_\gamma$
for one dimensional projection operators $P_\psi = \vert \psi
\rangle \langle \psi \vert$ on ${\mathcal{H}}_1
\otimes {\mathcal{H}}_2$ in terms of the coefficients in the
Schmidt representation of $\vert \psi \rangle$. In this section we
make extensive use of the Dirac bra-ket notation.
\begin{prop}
Let $\vert \psi \rangle \in {\mathcal{H}}_1 \otimes
{\mathcal{H}}_2$ be a unit vector and
$\vert \psi \rangle = \sum_i \sqrt{p_i}
\vert \phi_i \rangle
\otimes \vert \chi_i \rangle$ its Schmidt representation, where
$\{ \vert \phi_i \rangle \}_i$ and $\{ \vert \chi_i \rangle \}_i$ are
orthonormal bases of ${\mathcal{H}}_1$ and ${\mathcal{H}}_2$
respectively and where $p_i \geq 0$ and $\sum_i p_i = 1$.
Let $P_\psi$ denote the one
dimensional projection operator onto the subspace spanned by $\vert \psi
\rangle$. Then \[ \Vert P_\psi \Vert_\gamma = \sum_{ij} \sqrt{p_i p_j}
= \left( \sum_i \sqrt{p_i} \right)^{2}. \] \label{p4}
\end{prop}
\emph{Proof}: Without loss of generality we assume that
${\mathcal{H}}_1 = {\mathcal{H}}_2$ which can always be achieved
by possibly suitably enlarging one of the two Hilbert spaces.
Further, we
identify ${\mathcal{H}}_1 = {\mathcal{H}}_2$ with
${\mathbb{C}}^n$, where $n = \dim {\mathcal{H}}_1$, i.e., we fix
an orthonormal basis in ${\mathcal{H}}_1$ which we identify with the
canonical real basis in ${\mathbb{C}}^n$. With respect to this
canonical real basis in ${\mathbb{C}}^n$ we can define complex
conjugates of elements of ${\mathcal{H}}_1$ and the complex conjugate
as well as the transpose of a linear operator
acting on ${\mathcal{H}}_1$.
From the Schmidt decomposition it follows that
\begin{equation} P_\psi = \vert \psi \rangle \langle \psi \vert = \sum_{ij}
\sqrt{p_i  p_j} \vert \phi_i \rangle \langle \phi_j \vert \otimes
\vert \chi_i \rangle \langle \chi_j \vert. \label{0815}
\end{equation} From the definition of $\Vert \cdot \Vert_\gamma$ it
is thus obvious that $\Vert P_\psi \Vert_\gamma \leq \sum_{ij} \sqrt{p_i
p_j}$. Now consider the Hilbert space ${\mathfrak{H}}$
of Hilbert-Schmidt operators
on ${\mathcal{H}}_1 \otimes {\mathcal{H}}_2$ equipped with the
Hilbert-Schmidt inner product $\langle f \vert g \rangle =
{\mathrm{Tr}}(f^\dagger g)$. Equation (\ref{0815}) induces an
operator ${\mathfrak{A}}_\psi$ on ${\mathfrak{H}}$ as follows.
Every element $\zeta$ in ${\mathfrak{H}}$ can be written $\zeta =
\sum_k x_k \otimes y_k$ where $x_k$ and $y_k$ are trace class
operators on ${\mathcal{H}}_1$ and ${\mathcal{H}}_2$ respectively.
Then ${\mathfrak{A}}_\psi$ is defined on $\zeta$ as
${\mathfrak{A}}_\psi (\zeta) := \sum_{ijk} \sqrt{p_i p_j}
\langle \chi^*_i \vert x_k \vert \chi^*_j \rangle
\vert \phi_i \rangle \langle \phi_j \vert \otimes y_k$ where
$\vert \chi_i^* \rangle$ denotes the complex conjugate of the vector
$\vert \chi_i \rangle$ with respect to the canonical real basis in
${\mathbb{C}}^n$.
Proposition 11.1.8 in \cite{KadisonR86} implies that
${\mathfrak{A}}_\psi (\zeta)$ is independent of the
representation of $\zeta$. Consider a representation
$P_\psi = \sum_{i=1}^r u_i \otimes v_i$ of $P_\psi$ as sum over
simple tensors. Denote the transpose of $v_i$ by $v_i^T$.
Then the operator defined by
\begin{equation}
{\mathcal{A}}_\psi (\zeta) := \sum_{i,k=1}^r
{\mathrm{Tr}}(v_i^T x_k) u_i \otimes y_k \label{huggies}
\end{equation} is equal to ${\mathfrak{A}}_\psi$ (by virtue of
Proposition 11.1.8 in \cite{KadisonR86}).
We denote the trace on
${\mathcal{T}}({\mathfrak{H}})$ by $\tau(\cdot)$. The operator
${\mathfrak{A}}_\psi$ is of trace class and the right hand side of
Equation \ref{0815} is the so-called polar representation
of ${\mathfrak{A}}_\psi$ which implies $\tau({\mathfrak{A}}_\psi) =
\sum_{ij} \sqrt{p_i p_j}$, see \cite{Schatten70}.
${\mathfrak{A}}_\psi$ admits also many
other representations ${\mathfrak{A}}_\psi = \sum_i f_i \otimes
g_i$ with families of operators $\{ f_i \}$ and $\{ g_i \}$ acting
on ${\mathcal{H}}_1$ and ${\mathcal{H}}_2$ respectively.
It is well known that \[ \tau({\mathfrak{A}}_\psi) = \inf \left\{
\sum_i \Vert f_i \Vert_2 \Vert g_i \Vert_2 \, \left\vert
{\mathfrak{A}}_\psi = \sum_i f_i \otimes g_i \right.
\right\} \leq \Vert P_\psi \Vert_\gamma, \] where $\Vert \cdot
\Vert_2$ denotes the Hilbert-Schmidt norm and where the latter
inequality follows from $\Vert z \Vert_2 \leq \Vert z \Vert_1$ and
from the fact that each decomposition of ${\mathfrak{A}}_\psi$
corresponds in an obvious one-to-one fashion
to a decomposition of $P_\psi$.
This proves the proposition. $\Box$
\begin{co} \label{au}
Let $\rho$ be a density operator on ${\mathcal{H}}_1
\otimes {\mathcal{H}}_2$, where ${\mathcal{H}}_1$ and
${\mathcal{H}}_2$ are finite dimensional Hilbert spaces.
If $\rho = \sum_{ij}
a_{ij} \vert \phi_i \rangle \langle \phi_j \vert \otimes
\vert \chi_i \rangle \langle \chi_j \vert$, then $\Vert \rho \Vert_\gamma
= \sum_{ij} \vert a_{ij} \vert.$
\end{co}
An immediate corollary of Proposition \ref{p4} is that a pure
state $\vert \psi \rangle  \in {\mathcal{H}}_1 \otimes
{\mathcal{H}}_2$ is separable if and only if $\Vert P_\psi \Vert_\gamma
=1$.
In \cite{Rudolph00} it has been proven that more generally
all separable density
matrices can be characterized by $\Vert \cdot \Vert_\gamma$.
\begin{theo} \label{t1}
Let ${\cal H}_1$ and ${\cal H}_2$ be finite dimensional Hilbert
spaces and $\varrho$ be a density operator on ${\cal H}_1
\otimes {\cal H}_2$. Then $\varrho$ is separable if and only if
$\Vert \varrho \Vert_\gamma = 1$. \end{theo}
In \cite{Rudolph00} it has been tentatively suggested that
$\Vert \cdot \Vert_\gamma$ can be considered as a quantitative measure of
entanglement. In the present work we substantiate this claim by
proving
\begin{prop} \label{p1} The function \[
E_\gamma(\sigma) := \Vert \sigma \Vert_\gamma
\log \Vert \sigma \Vert_\gamma  \] satisfies
the criteria (E0) - (E4) for entanglement measures. \end{prop}
\emph{Proof}: (E1) is an immediate consequence of Theorem
\ref{t1} and (E0) and (E2) are clear. (E3): Let $T$ be an
operation composed
of local operations, and classical communication. As we have
argued above every such $T$ can be realized as a sequence of
operations of the form $T_1 \otimes T_2$ where $T_1$ and $T_2$ are
local operations on system 1 and 2 respectively.
We show that $\Vert (T_1 \otimes T_2) (\sigma) \Vert_\gamma \leq \Vert
\sigma \Vert_\gamma$.
By linearity of $T_1 \otimes T_2$ every decomposition of $\sigma$
into finite sums of simple tensors $\sigma = \sum_{i=1}^r x_i \otimes
y_i$, where $x_i$ and $y_i$ are trace class operators on
${\mathcal{H}}_1$ and ${\mathcal{H}}_2$ respectively,
induces a decomposition of $(T_1 \otimes T_2) (\sigma)$ into a sum of
simple tensors $(T_1 \otimes T_2) (\sigma) =
\sum_{i = 1}^r T_1 (x_i) \otimes
T_2 (y_i)$. Thus
\begin{eqnarray*} \Vert (T_1 \otimes T_2) (\sigma) \Vert_\gamma
& = & \inf \left\{ \sum_{i=1}^r \Vert X_i \Vert_1 \Vert Y_i \Vert_1 \,
\left\vert \, (T_1 \otimes T_2) (\sigma) =
\sum_{i=1}^r X_i \otimes Y_i \right. \right\}
\\ & \leq & \inf \left\{ \sum_{i=1}^r \Vert T_1(x_i) \Vert_1 \Vert
T_2(y_i) \Vert_1 \, \left\vert
\, \sigma = \sum_{i=1}^r x_i \otimes y_i \right. \right\} \\
& \leq & \Vert T_1 \Vert \, \Vert T_2 \Vert \inf
\left\{ \sum_{i=1}^r \Vert x_i \Vert_1 \Vert y_i \Vert_1 \, \left\vert
\, \sigma = \sum_{i=1}^r x_i \otimes y_i \right. \right\} \\
& \leq & \inf
\left\{ \sum_{i=1}^r \Vert x_i \Vert_1 \Vert y_i \Vert_1 \, \left\vert
\, \sigma = \sum_{i=1}^r x_i \otimes y_i \right. \right\} \\
& = & \Vert \sigma \Vert_\gamma,
\end{eqnarray*} where we have used that both $T_1$ and $T_2$ are
bounded maps on the space of trace class operators on
${\mathcal{H}}_1$ and ${\mathcal{H}}_2$ respectively
and that \[ \Vert T_i \Vert = \sup \{ {\mathrm{Tr}}(T_i (\rho)) \,
\vert \, \rho \in {\mathcal{T}}({\mathcal{H}}_i), \rho \geq 0 \, \,
\mathrm{ and }
\, \, {\mathrm{Tr}}(\rho) = 1 \} \leq 1, \] see, e.g., Lemma 2.2.1 in
\cite{Davies76}. (E3) follows from the fact that
$[1, \infty[ \ni s \mapsto s \log s$ is monotone.
Finally, (E4) follows from the facts that $\Vert \cdot \Vert_\gamma$ is
subadditive and that $[1, \infty[ \ni s
\mapsto s \log s$ is monotone and convex. $\Box$ \\
\begin{rema}
It follows from the proof of Proposition \ref{p1} that if $f$ is a
convex, monotonously increasing function on $[1, \infty[$ with $f(1) =
0$, then \[ E_f(\sigma) :=
f( \Vert \sigma \Vert_\gamma) \] is an entanglement measure
satisfying the requirements (E0) - (E4).
A possible choice is $f_1(x) = x-1$ leading to the entanglement measure
$E_{f_1}(\sigma) = \Vert \sigma \Vert_\gamma -1$. This shows that
indeed (as claimed in \cite{Rudolph00}) the function
$\Vert \sigma \Vert_\gamma -1$ is an entanglement measure on the
space of density operators. Other possible choices for $f$ are
$f_2(x) = x \ln x -x +1$, $f_3(x) = e^{a(x-1)}, a>0$ and so forth.
\label{beepbeep} \end{rema} \begin{co}
The entanglement measures constructed in Remark \ref{beepbeep}
(including the measure $E_\gamma$ from
Proposition \ref{p1})
satisfy that $E_f(\sigma) =0$ if and only if $\sigma$ is
separable. \end{co}
\emph{Proof}:
This is an immediate consequence of Theorem \ref{t1}. $\Box$

\begin{prop}
Let $T_1$ and $T_2$ be two
trace-preserving L\"uders-von Neumann operations on finite
dimensional Hilbert spaces
${\mathcal{H}}_1$ and ${\mathcal{H}}_2$ respectively and
let $T_L = T_1 \otimes T_2$ denote the corresponding
L\"uders-von Neumann operation
acting locally on ${\mathcal{H}}_1 \otimes {\mathcal{H}}_2$.
Let $T_1(\sigma_1) = \sum_i P_i \sigma_1 P_i$ and $T_2(\sigma_2) =
\sum_j Q_j \sigma_2 Q_j$ be Choi-Kraus representations of $T_1$ and $T_2$
respectively
in terms of families $\{ P_i \}$ and $\{ Q_j \}$ of, respectively,
mutually orthogonal
projection operators. Then the entanglement measure
$E_{f_1} = \Vert \cdot \Vert_\gamma -1$
as in Remark \ref{beepbeep} satisfies \[
\sum_{ij} p_{ij} \left(\Vert \sigma_{ij} \Vert_\gamma - 1 \right)
\leq \Vert \sigma \Vert_\gamma -1 \]
where $p_{ij} := {\mathrm{Tr}}((P_i \otimes Q_j) \sigma (P_i \otimes
Q_j))$ and $\sigma_{ij} = \frac{(P_i \otimes Q_j) \sigma (P_i
\otimes Q_j)}{p_{ij}}$ and where $\sigma$ is a density
operator on ${\mathcal{H}}_1 \otimes {\mathcal{H}}_2$.
\label{p8} \end{prop}
\emph{Proof}:
Let $P$ and $P'$ be orthogonal projection operators. Then
$\Vert P x P + P' y P' \Vert_1 = \Vert P x P \Vert_1 + \Vert P' y P'
\Vert_1$ for all operators $x,y$. This follows from considering the
spectral resolutions of $P x P$ and $P' y P'$. Hence
$\sum_i \Vert P_i x_k P_i \Vert_1
= \Vert \sum_i P_i x_k P_i \Vert_1
\leq \Vert x_k \Vert_1$. A similar argument shows that $\sum_j \Vert Q_j \tilde{z} Q_j \Vert_1 \leq
\Vert \tilde{z} \Vert_1$ for all $\tilde{z} \in
{\mathcal{B}}({\mathcal{H}}_2)$. Hence
\begin{eqnarray*}
\sum_{ij} p_{ij} \left( \Vert \sigma_{ij} \Vert_\gamma -1 \right)
& \leq & \inf \left\{ \sum_{ijk} \Vert P_i x_k P_i \Vert_1 \Vert
Q_j y_k Q_j \Vert_1 \, \left\vert \sigma = \sum_k x_k \otimes y_k
\right. \right\} - 1 \\ & \leq & \inf \left\{
\sum_{k} \Vert x_k \Vert_1 \Vert
y_k \Vert_1 \, \left\vert \, \sigma = \sum_k x_k \otimes y_k
\right. \right\} -1 \\ & = & \Vert \sigma \Vert_\gamma -1. \Box
\end{eqnarray*}

It is known that some physically interesting entanglement measures
coincide with the von Neumann reduced entropy on pure states, for
instance the relative entropy of entanglement \cite{PlenioV98}.
However, it follows immediately from Proposition
\ref{p4} that $E_\gamma$ does not coincide with the von Neumann
reduced entropy on pure states: it follows from \cite{PlenioV98}
that the entropy of entanglement for a pure state of the form $
\vert \phi \rangle = \alpha
\vert 00 \rangle + \beta \vert 11 \rangle$ is equal to
$-\vert \alpha \vert^2 \ln \vert \alpha \vert^2
-\vert \beta \vert^2 \ln \vert \beta \vert^2$, whereas it follows
from Proposition \ref{p4} that $E_\gamma\left( \vert \phi \rangle
\langle \phi \vert \right) = 2 \left( \vert \alpha \vert + \vert \beta
\vert \right)^2 \ln \left( \vert \alpha \vert + \vert \beta \vert \right).$
Therefore we have explicitly constructed an
entanglement measure satisfying a physically
reasonable set of requirements
which is not equal to the von Neumann reduced entropy on pure states.
We have proven
\begin{prop} \label{17}
$E_\gamma$ and $S_{\mathrm{vN}}$ do not coincide on pure states.
\end{prop}
In \cite{DonaldHR01} necessary and sufficient conditions for an
entanglement measure to coincide with $S_{\mathrm{vN}}$ on pure states
were
derived. It is easy to see that, e.g., $E_\gamma$ does not satisfy
the additivity condition (P4) considered in \cite{DonaldHR01}.
\section{Higher Tensor Product Hilbert Spaces}
So far we restricted ourselves to tensor
product Hilbert spaces of two finite dimensional Hilbert spaces.
It is straightforward, however, to generalize our results to the
situation of tensor products of more than two, but at most
finitely many, finite dimensional Hilbert spaces.
To this end consider the tensor product ${\mathcal{H}} = $ \linebreak[3]
${\mathcal{H}}_1 \otimes
\cdots \otimes {\mathcal{H}}_n$ of $n$ finite dimensional Hilbert
spaces ${\mathcal{H}}_1, \cdots, {\mathcal{H}}_n$.
The obvious generalization of the definition of $\Vert \cdot
\Vert_\gamma$ is
\begin{equation}
\Vert t \Vert_\gamma^{(n)} := \inf \left\{ \sum_{i=1}^r
\left\Vert u^{(1)}_i \right\Vert_1 \cdots \left\Vert
{u}^{(n)}_i \right\Vert_1 \, \left\vert \, t = \sum_{i=1}^r u^{(1)}_i
\otimes \cdots \otimes {u}^{(n)}_i \right. \right\},
\end{equation}
where $t$ is a trace class operator on $\mathcal{H}$.

It is straightforward to generalize the main result of \cite{Rudolph00} to
$n$-fold tensor product Hilbert spaces ${\mathcal{H}} =$ ${\mathcal{H}}_1
\otimes \cdots \otimes {\mathcal{H}}_n$
\begin{de}
Let ${\cal H}_1, \cdots, {\cal H}_n$ be Hilbert spaces of
arbitrary dimension. A density operator $\varrho$ on the tensor product
${\cal H}_1 \otimes \cdots
\otimes {\cal H}_n$ is called \emph{disentangled} or
\emph{separable} (with respect to ${\cal H}_1, \cdots, {\cal
H}_n$) if there exist a
family $\left\{ \omega_{i} \right\}$ of positive real numbers, and
families $\left\{ \rho^{(k)}_i \right\}$ of density operators on
${\cal H}_k$ respectively, where $1 \leq k \leq n$,
such that
\begin{equation} \label{esep}
\varrho = \sum_{i} \omega_{i} \rho^{(1)}_i \otimes \cdots
\otimes \rho^{(n)}_i,
\end{equation}
where the sum converges in trace class norm.
\end{de} \begin{theo} \label{tsep}
Let ${\cal H}_1, \cdots, {\cal H}_n$ be finite dimensional Hilbert
spaces and $\varrho$ be a density operator on ${\mathcal{H}} =$
${\cal H}_1 \otimes
\cdots \otimes {\cal H}_n$. Then $\varrho$ is separable if and only if
$\Vert \varrho \Vert^{(n)}_\gamma = 1$. \end{theo}
We now consider the situation that ${\mathcal{H}}$ is the $m$-fold
tensor product of ${\mathfrak{H}}_1 \otimes {\mathfrak{H}}_2$ with
two finite dimensional Hilbert spaces ${\mathfrak{H}}_1$ and
${\mathfrak{H}}_2$. The functional $E_\gamma$ from Proposition
\ref{p1} admits an obvious extension
\begin{equation}
\label{12}
E_\gamma(\sigma) := \Vert \sigma \Vert_\gamma^{(n)} \ln \Vert
\sigma \Vert_\gamma^{(n)}
\end{equation} for all trace class operators $\sigma$ on
${\mathcal{H}}$.
\begin{prop}
The functional defined by Equation (\ref{12}) satisfies the
criteria (E0)-(E4) for entanglement measures.
\end{prop}
\section{Conclusion}
To conclude, in this paper
we have introduced a new class of entanglement
measures on the space of density operators on tensor product
Hilbert spaces. Our entanglement measures are based on the
greatest cross norm $\Vert \cdot \Vert_\gamma$
on the set of trace class operators on the
tensor product Hilbert space.
We showed that our entanglement measures satisfy a number of
physically desirable requirements, in particular that they do not
increase under local quantum operations. \\

\noindent\textbf{Acknowledgements} \\
I thank Matthew J.~Donald, Micha\l \/ Horodecki and Michael Wolf
for their comments on a previous version of this paper and
Armin Uhlmann for pointing out Corollary \ref{au} to me.

\end{document}